\documentclass[%
 reprint,superscriptaddress,
 amsmath,amssymb,
 aps,prb,floatfix,longbibliography,nofootinbib]{revtex4-2}
\usepackage[normalem]{ulem}
\usepackage{graphicx}
\usepackage{dcolumn}
\usepackage{bm}
\usepackage[colorlinks=true, linkcolor=red, breaklinks=true, citecolor=blue, urlcolor=red]{hyperref}
\usepackage{silence}
\usepackage{algorithm}
\usepackage{amsfonts}
\usepackage{booktabs}
\usepackage{siunitx}
\usepackage{ulem}
\usepackage{tabularray}
\usepackage{algpseudocode}

\usepackage{braket}

\usepackage[status=draft,inline,nomargin]{fixme}
\fxusetheme{color}
\FXRegisterAuthor{au}{aku}{\color{magenta}AU}

\usepackage{xcolor}

\newcommand{\changes}[1]{\textcolor{black}{#1}}
\begin{document}

\preprint{APS/123-QED}

\title{Finite-size scaling on the torus with periodic projected entangled-pair states}

\author{Gleb Fedorovich}
\email{gleb.fedorovich@ugent.be}
\author{Lukas Devos}
\author{Jutho Haegeman}
\affiliation{Department of Physics and Astronomy, Ghent University, Krijgslaan 281, 9000 Gent, Belgium}
\author{Laurens Vanderstraeten}
\affiliation{Center for Nonlinear Phenomena and Complex Systems, Universit\'e Libre de Bruxelles, Belgium}
\author{Frank Verstraete}
\affiliation{Department of Physics and Astronomy, Ghent University, Krijgslaan 281, 9000 Gent, Belgium}
\affiliation{Department of Applied Mathematics and Theoretical Physics, University of Cambridge,\\ Wilberforce Road, Cambridge, CB3 0WA, United Kingdom}
\author{Atsushi Ueda}
\affiliation{Department of Physics and Astronomy, Ghent University, Krijgslaan 281, 9000 Gent, Belgium}

\date{\today}

\begin{abstract}
An efficient algorithm is constructed for contracting two-dimensional tensor networks under periodic boundary conditions. The central ingredient is a novel renormalization step that scales linearly with system size, i.e., from \(L \to L+1\). The numerical accuracy is comparable to state-of-the-art tensor network methods, while giving access to much more data points, and at a lower computational cost. Combining this contraction routine with the use of automatic differentiation, we arrive at an efficient algorithm for optimizing fully translation invariant projected entangled-pair states on the torus. Our benchmarks show that this method yields finite-size energy results that are comparable to those from quantum Monte Carlo simulations. When combined with field-theoretical scaling techniques, our approach enables accurate estimates of critical properties for two-dimensional quantum lattice systems.
\end{abstract}

\maketitle

\section{Introduction}

Quantum many-body systems at criticality exhibit exciting emergent phenomena. The unifying picture for understanding quantum criticality is given by the renormalization group: the low-energy or long-distance properties of critical systems are described by field theories. Originally, it was assumed that these field theories are always related to fluctuations of some local order parameters that characterize the phase transition, but nowadays several instances are known of exotic quantum critical points that defy this standard picture.

In this context, the role of numerical methods is crucial for establishing whether these more exotic instances of quantum criticality can occur in microscopic models. Finite computational resources, however, can never reproduce the nonanalytic properties of an infinite system at criticality. In numerical approaches such as exact diagonalization or quantum Monte Carlo, it is the finite size of the system that leads to a regularization of the critical signatures. Fortunately, the theory of finite-size scaling was developed~\cite{Fisher1972, Binder:1981sa, Brezin1982, Cardy1988}: the critical properties can be determined numerically from the scaling of different quantities as the system size is increased. With the formulation of powerful scaling hypotheses, this approach has led to many successes in revealing the universal critical properties in many microscopic models, or, on the contrary, showing that certain models in fact do not obey the expected scaling properties.

Another numerical approach for studying quantum critical systems is using tensor networks as variational wavefunctions for approximating their ground states \cite{Xiang_2023}. In contrast to finite-size methods, tensor networks can be formulated directly in the thermodynamic limit when some kind of translation invariance is present in the system, but they are limited in the amount of entanglement they can capture -- this limitation is typically expressed as the bond dimension of the tensor network, which can be systematically increased to reach more precise results. In analogy to finite-size scaling, a theory of finite-entanglement scaling~\cite{Nishino1996, tagliacozzo2008scaling, pollmann2009theory, pirvu2012matrix} is currently under construction to extract the critical properties of a given many-body system systematically. For one-dimensional (1-D) systems, methods based on the formalism of matrix product states (MPS) are extremely efficient for obtaining numerical results at very high bond dimensions \cite{SCHOLLWOCK201196}, and a number of tools \cite{rams2018, Pillay2019,vanhecke2019, Eberharter2023, Huang2024, schneider2024} are available for extracting critical properties for generic models. 

In a parallel development, tensor networks have also been used to encode the ideas of the renormalization group in a more explicit sense. For a 1-D quantum system, the multiscale entanglement renormalization ansatz (MERA) \cite{Vidal2007, Vidal2008} was proposed to represent the ground states of critical models. Although this approach is less efficient than the MPS-based algorithms, the scaling properties of critical ground states are naturally represented and precise scaling dimensions were determined for some critical models \cite{Giovanetti2008, Pfeifer2009}. In the context of statistical mechanics, similar ideas have led to different real-space renormalization schemes such as the tensor renormalization group (TRG) \cite{Levin2007, Gu2008, Xie2009, Xie2012,RevModPhys.94.025005,PhysRevE.89.013308} and tensor network renormalization (TNR) \cite{Evenbly2015, Evenbly2017, Yang2017,BalPRL}, which capture the scaling properties of critical models in two dimensions and beyond.
\begin{figure*}[tb]
    \centering
    \includegraphics[width=2\columnwidth]{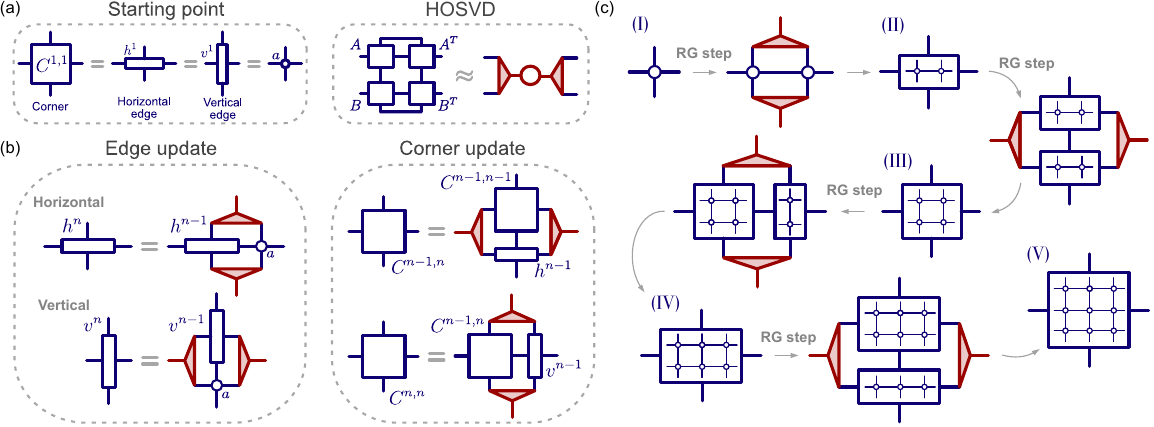}
    \caption{
Schematic illustrations of the coarse-graining procedure in PTMRG. The coarse-grained tensors consist of four types of tensors: $a$, $h^n$, $v^n$, and $C^{n,m}$. In the context of a PEPS simulation, $a$ represents the inner product of the two local PEPS tensors, whereas it is the local tensor encoding the Boltzmann weights in the context of a partition function. Tensors $h^n$, $v^n$ and $C^{n,m}$ respectively represent coarse-grained tensor of $n\times1$, $1\times m$, and $n\times m$ tensor network of $a$. (a) All initial tensors are set to be $a$. The isometries involved in RG steps are obtained via HOSVD. (b) Throughout the RG, $h$ and $v$ are iteratively combined with $a$ to expand the system size by one. Similarly, $C$ is combined with $h$ and $v$ when we expand the system in the vertical and horizontal directions, respectively. (c) An example of $3$ by $3$ PEPS tensor network contracted via PTMRG scheme that alternately coarse-grains the tensor in the vertical and horizontal direction twice.}
    \label{fig:scheme_1}
\end{figure*}

For two-dimensional (2-D) quantum systems, the formalism of projected entangled pair states (PEPS)~\cite{PhysRevA.70.060302, verstraete2004} and tree tensor networks (TTN) \cite{Shi2006, Tagliacozzo2009, Murg2010} can be used efficiently to represent correlated ground states. Again, PEPS can be formulated and manipulated directly for infinite system sizes \cite{Jordan2008Dec}, but extracting critical properties is less straightforward. On the one hand, the numerical methods are computationally more demanding, so the accessible range of bond dimensions is smaller than in the 1-D MPS case. On the other hand, the theory of finite-entanglement scaling for PEPS~\cite{Rader2018, PhysRevX.8.031031, Vanhecke2022, PhysRevLett.131.266202} is still in full development.

Given this situation, it makes sense to investigate whether the use of variational tensor network methods for performing finite-size scaling could lead to a complementary approach to studying quantum critical systems. The main objection concerns the fact that finite-size scaling is performed preferably on systems with periodic boundary conditions, which are a lot more demanding to simulate with tensor networks than systems with open boundary conditions. Nonetheless, several works~\cite{Porras2006, Pippan2010, Rossini2011, Pirvu2011, Draxler2017, Zou2018, Zou2020, Zou2020b, Zou2021, VanDamme2021} have shown that MPS can be used efficiently for periodic systems to extract critical properties. In particular, in Refs.~\cite{Pirvu2011, Zou2018}, it was shown that uniform MPS on a periodic system can be optimized efficiently. In Ref.~\cite{Dong:2024hyf}, a periodic PEPS algorithm was constructed by using a PEPS with open boundary conditions and taking an extensive superposition of translated copies to restore the translation symmetry of the system. By using sampling techniques and stochastic optimization, this ansatz was efficiently optimized using high-performance and parallelized computing. 

In this work, we introduce an algorithm for directly optimizing a single periodic PEPS wavefunction. This approach implements the translation invariance of the periodic geometry in the most natural way, thereby requiring significantly fewer variational parameters for the same accuracy. Our algorithm is based on an efficient contraction routine of periodic 2-D tensor networks, which combines ideas from the real-space renormalization ideas that have been used for simulating 2-D stat-mech models, such as TRG/TNR and the corner transfer matrix renormalization group (CTMRG)~\cite{Baxter1968, Baxter1978, nishino1996corner, Nishino_1997}. We then combine this contraction routine with the use of automatic differentiation (AD) to variationally optimize periodic PEPS wavefunctions.

\section{Contraction of periodic 2D tensor networks}

The backbone of any PEPS algorithm is an efficient routine to accurately approximate expectation values by contracting the resulting tensor network. For infinite PEPS, boundary MPS methods \cite{verstraete2004, Fishman2018Dec, Vanderstraeten2022} and CMTRG variants \cite{Baxter1968, Baxter1978, nishino1996corner, Nishino_1997, Orus2009, Corboz2010, Corboz2014} are the most viable options, but these cannot be readily applied in the periodic case. Real-space RG methods such as TRG and TNR can be used to efficiently contract periodic systems~\cite{PhysRevB.100.035449, PhysRevB.102.054432, PhysRevB.105.L060402, Kadoh:2019kqk, Akiyama:2024ush, Bal2017, Hauru2018, homma2023nuclear}, but typically implement scale transformations of the lattice so that system sizes have to come in logarithmic spacings if translation invariance is to be respected after every coarse-graining step. In this work, therefore, we propose a periodic transfer \changes{matrix} renormalization group (PTMRG) method, which combines ideas from the CTMRG and TRG approaches.

\subsection{Periodic transfer matrix renormalization group (PTMRG)\label{sec:PTMRG}}

The main steps of the proposed PTMRG scheme are illustrated in Fig. \ref{fig:scheme_1}. Given a homogeneous $L \times L$ network of contracted four-leg tensors $a$, we first introduce a coarse-grained tensor $C^{n,m}$ and a vertical (horizontal) edge tensor $v^{n}\ (h^{n})$, where the superscripts indicate the system size associated with each tensor. In other words, $C^{n,m}$ describes a system with $n\times m$ sites in total, while $v^{n}\ (h^{n})$ represents a vertical (horizontal) $n$-site chain containing $a$ tensors. 

Initializing $C^{1,1} = h^1 = v^1 = a$, we develop the update rules that are presented in Fig.~\ref{fig:scheme_1}(b). More precisely, following the standard HOSVD approach~\cite{DeLathauwer2006Jul, PhysRevB.100.035449}, we obtain isometries based on $C^{n-1,n-1}$ and $h^{n-1}$ environment tensors, and perform a single horizontal sweep to get $C^{n-1,n}$ and $h^{n}$. As usual for most TRG schemes,  the isometries guarantee to avoid exponentially growing bond dimension by introducing a threshold for the truncation dimension $\chi$. In the same way we perform a vertical sweep to get $C^{n,n}$ and $v^{n}$ from $C^{n-1,n}$ and $v^{n-1}$. By sequentially applying horizontal and vertical transformations $L$ times, the final coarse-grained tensor $C^{L,L}$ represents the $L\times L$ system. Finally, since we are mostly interested in studying finite systems with periodic boundary conditions (on a torus), one can straightforwardly contract the top-bottom and left-right legs of the coarse-grained environment tensors. 

Regarding the CPU time required to perform a single RG contraction, the proposed PTMRG scheme has a reduced scaling by using at each renormalization step a given local four-leg tensor $a$ with fixed bond dimension $D$. The computational complexity of TRG and HOTRG schemes scale as $O(\chi^6)$~\cite{Levin2007} and $O(\chi^7)$~\cite{PhysRevB.100.035449} respectively, while the PTMRG algorithm scales as $O(\chi^5)$ \changes{-- we refer to Appendix~\ref{appendix:cost_PTMRG} for a detailed analysis of the scaling of the algorithm}. We confirm this scaling behavior in the following subsection while benchmarking the evaluation of the partition function of the 2-D classical Ising model.

\subsection{Application to 2-D statistical mechanics}

To demonstrate the performance of the PTMRG contraction scheme, we consider the classical Ising model on a square lattice, 
\begin{equation}
    E_{\text{Ising}}(\left\{\sigma_i\right\}) = -\sum_{\braket{ij}}\sigma_i\sigma_j, \qquad \sigma_i=\pm 1
\end{equation}
with PBC at the critical temperature $T_c = 2/\ln(1+\sqrt{2})$. When computing the statistical model, we aim to evaluate the partition function with PBC, denoted as ${Z}(L,L)$. For the classical Ising model on a square lattice, this is represented by $L \times L$ contractions of the four-leg local tensor $T^{(1)}_{ijkl}$, as described in Ref.~\cite{PhysRevD.88.056005}:
\begin{align}
    T^{(1)}_{ijkl} = 2\cosh^2(\beta)f_if_jf_kf_l\delta_{\textrm{mod}[i+j+k+l,2],0}\nonumber,
\end{align}
where $\beta$ is the inverse temperature, and $f_n = \sqrt{\tanh(\beta)}^{\ n}$. The indices $i,j,k,l$ (left, top, right, down) take values of 0 or 1. Contracting these tensors corresponds to summing over spin configurations, and performing this contraction on a $L \times L$ lattice enables us to compute the entire partition function. Specifically, we use $C^{L,L}$ from PTMRG to compute the partition function starting from the initial tensor $a = T^{(1)}_{ijkl}$.

\begin{figure}[tb]
    \centering
    \includegraphics[width=1\columnwidth]{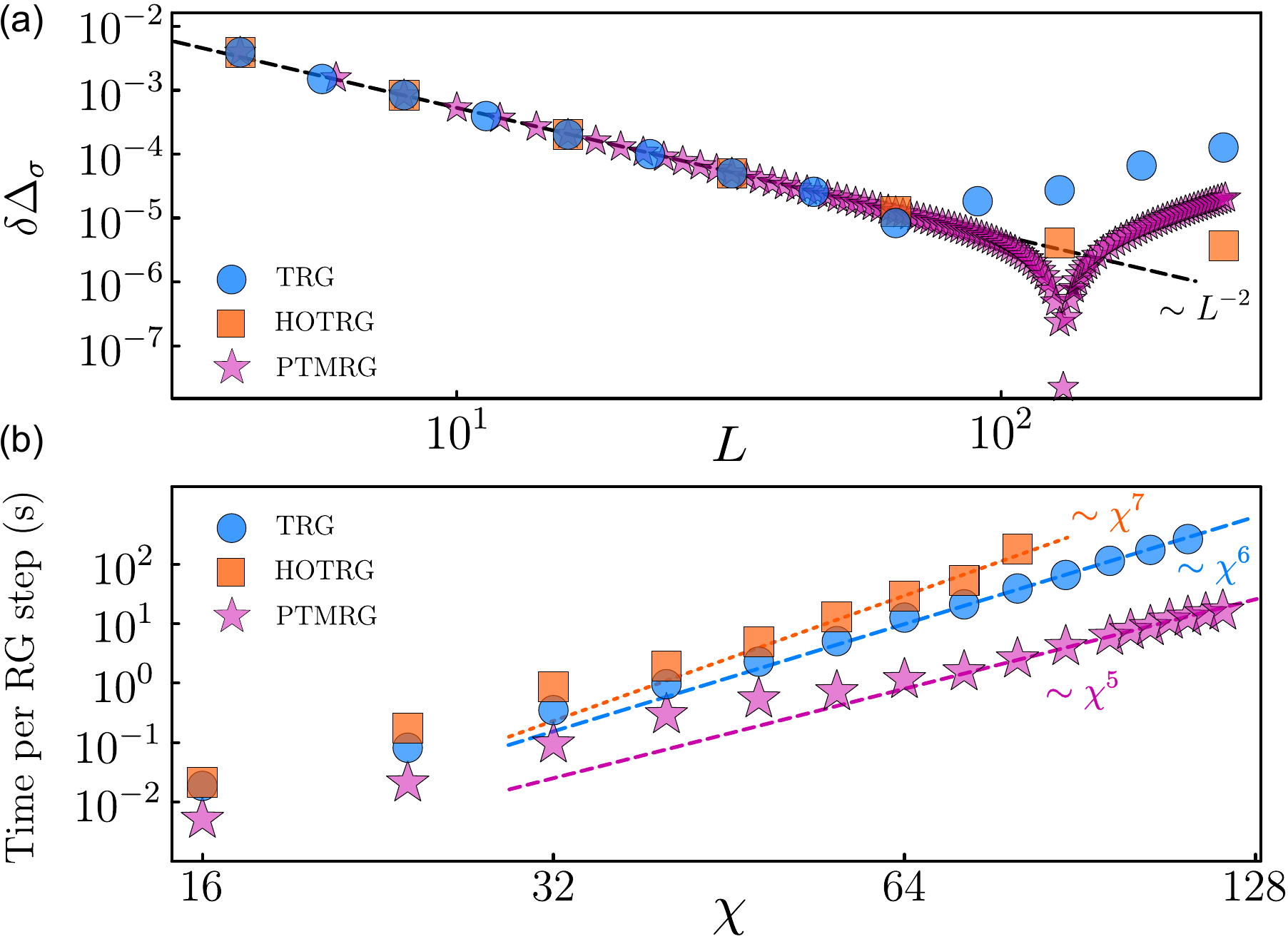}
    \caption{(a) Finite-size effect $\delta \Delta_\sigma(L)$ computed from TRG($\chi=64$), HOTRG($\chi=56$), and PTMRG($\chi=112)$. The theoretical scaling $\propto 1/L^2$ is shown by a black dashed line. (b) The average computational time per RG step for the critical Ising model. The dashed lines show the expected asymptotic scalings with respect to the truncation dimension $\chi$.}
    \label{fig:stat_mech}
\end{figure}

Figure~\ref{fig:stat_mech}(a) shows the computational time required to perform one update step for TRG, HOTRG, and PTMRG~\footnote{We calculate the average time consumed per step after omitting the first few steps where the bond dimension is still growing to its maximal value.}. When using the same bond dimension $\chi$, PTMRG is faster by one order of magnitude. Consequently, more data points can be collected in the same computational time.

In addition to its computational efficiency, PTMRG also demonstrates high accuracy. TRG/TNR algorithms are known to introduce numerical errors due to the bond dimension truncation~\cite{Levin2007,Evenbly2015,Evenbly2017,Yang2017,Bal2017,Hauru2018,PhysRevB.104.165132,PhysRevB.108.024413,PhysRevB.107.205123}, with these errors being most severe in critical systems with a diverging correlation length. This challenging scenario can be used to estimate the accuracy of the algorithm. At criticality, the transfer matrix, obtained by contracting two horizontal legs of $C^{L,L}$, has a universal spectrum $\Lambda_n$~\footnote{To improve accuracy, we construct a $2\times1$ transfer matrix from $C^{L,L}$ when computing Fig.~\ref{fig:stat_mech}(a).}. This $\Lambda_n$ corresponds to the scaling dimension $\Delta_n$ of the conformal field theory (CFT) as
\begin{align}
    \Lambda_n = A \exp(-2\pi \Delta_n),
\end{align}
where the eigenvalues are ordered in descending order, and $A$ is a nonuniversal constant~\cite{Cardy_1984,Cardy_1986,PhysRevB.80.155131}. For unitary CFTs, $\Delta_0$ is zero, so we can obtain the scaling dimension by taking the ratios of the eigenvalues as 
\begin{align}
    \Delta_n = \frac{1}{2\pi}\ln(\Lambda_0/\Lambda_n).\nonumber
\end{align}
This universal spectrum can be used to gauge the numerical error because the deviation from the universal value is attributed to the numerical error as discussed in Ref.~\cite{PhysRevB.108.024413}. In particular, the deviation in the second leading eigenvalue is known to capture this effect well. We define this deviation as
\begin{align}
    \delta \Delta_\sigma = |\Delta_\sigma-\Delta_\sigma(L)|,
\end{align}
where $\Delta_\sigma(L) = \frac{1}{2\pi} \ln(\Lambda_0 / \Lambda_1)$ and $\Delta_\sigma = \frac{1}{8}$ is the scaling dimension for the corresponding state. In addition to the numerical errors, $\Delta_\sigma$ contains a finite-size effect that decays as $\propto L^{-2}$ due to the leading irrelevant operator. Thus, the numerical error should manifest as a deviation of $\delta \Delta_\sigma$ from this $\sim L^{-2}$ line.

Figure~\ref{fig:stat_mech}(b) displays $\delta \Delta_\sigma$ obtained from TRG ($\chi=64$), HOTRG ($\chi=56$), and PTMRG ($\chi=112$). We choose these bond dimensions so that the computational time per step is the same, as can be read off from $(a)$. First, we find that the TRG data points quickly drift away from the expected finite-size scaling $\sim L^{-2}$ denoted with a dotted line around $L=64$, allowing us to use only eight data points as reliable data. Similarly, there are only five reliable HOTRG data points up to $L=64$ before the deviation at $L=128$. However, the data points of PTMRG exhibit correct scaling up to $L\sim 100$ with approximately fifty data points. From this point of view, PTMRG has a significant advantage over TRG and HOTRG, allowing us to obtain more data points before truncation errors become significant. We expect this aspect to prove valuable for finite-size scaling in statistical mechanics models, providing more reliable results in TRG schemes.

\section{Periodic PEPS algorithm}

Given an efficient method for contracting periodic 2-D tensor networks, we can now develop our periodic PEPS algorithm. For infinite PEPS, the original algorithms for optimizing the local tensors were based on applying imaginary time evolution via the Trotter-Suzuki decomposition to a random initial state, using varying levels of accuracy and sophistication known as simple update~\cite{Jiang2008Aug}, full update~\cite{Jordan2008Dec}, cluster update~\cite{Wang2011ClusterUF} and fast-full update~\cite{PhysRevB.92.035142}. More recently, progress was made using various direct energy minimization approaches based on DMRG-like optimisation~\cite{Corboz2016Jul} as well as gradient-based numerical optimization methods~\cite{Vanderstraeten2016Oct}. These more recent strategies outperform the methods based on imaginary time evolution in terms of their ability to find the optimal state at a given bond dimension. Furthermore, the gradient-based methods have become very efficient by incorporating the use of automatic differentiation (AD)~\cite{baydin2018automatic}: given an efficient contraction routine for evaluating the variational energy, one can simply apply AD to compute the gradient \cite{PhysRevX.9.031041} and optimize the variational energy. 

However, all of these developments are centered around uniform PEPS in the thermodynamic limit. Optimization of finite PEPS \cite{Lubasch2014, PhysRevB.107.165112} has proven to be considerably more difficult, although the sampling and stochastic optimization of generic PEPS \cite{Liu2017, Vieijra2021, Dong:2024hyf} and TEBD or DMRG like optimizations for the restricted set of isometric PEPS \cite{PhysRevLett.124.037201, Lin2022} have been carried out successfully. All of these approaches are formulated for systems with open boundary conditions, where all the tensors have to be chosen differently. In the absence of a canonical form for PEPS, this makes the optimization very costly.

\begin{figure}[tb]
    \centering
    \includegraphics[width=\columnwidth]{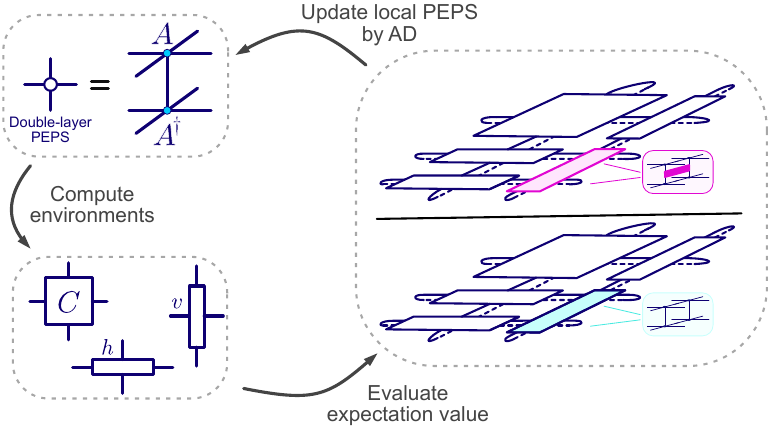}
    \caption{Schematic picture of the periodic PEPS algorithm. We minimize the expectation value of the local Hamiltonian with respect to the local PEPS tensor, as shown in Eq.~\eqref{eq:variational_GS}. The $n\times n$ contraction of $a$ tensors is replaced by the coarse-grained tensors to maintain the numerical tractability. For the numerator of Eq.~\eqref{eq:variational_GS}, two $a$ are replaced by the two-site impurity denoted by a pink rectangular tensor. Once the evaluation is done, AD minimizes this cost function using gradient-based methods.}
    \label{fig:PTMRG_scheme}
\end{figure}

On the torus, the situation is very different. Because of translation invariance, we can choose a uniform PEPS parametrization in terms of a single local tensor $A$ with bond dimension $D$, similar to the case of infinite PEPS. Additionally, we can impose other symmetries on the PEPS tensor such as reflection and rotation symmetries. The contraction of a periodic PEPS can be performed using the PTMRG approach of the previous section for a double-layer tensor network, whereas the evaluation of a local expectation value is performed by a single extension (see Fig.~\ref{fig:PTMRG_scheme}). The optimization problem that we want to solve is then given by
\begin{align}
    \epsilon_0(A) = \min_{A}\frac{\langle\Psi(A)|h|\Psi(A)\rangle}{\langle\Psi(A)|\Psi(A)\rangle}\label{eq:variational_GS},
\end{align}
where $|\Psi(A)\rangle$ denotes the PEPS wavefunction that is generated by the tensor $A$, and $h$ is a nearest-neighbour term in the hamiltonian. Note that we have to evaluate only a single term since all the other terms in the Hamiltonian are identical due to the translation or rotation symmetry of the PEPS. This optimization problem is now solved using the automatic differentiation (AD) approach.

\par The proposed algorithm is shown in Fig.~\ref{fig:PTMRG_scheme} and contains the following main steps:
\begin{algorithmic}[1]
    \State Merge the given local $A$-PEPS and $A^\dagger$-PEPSs into an $a$-tensor
    \State Using the PTMRG scheme, obtain environment tensors $C^{L-1,L-2}$ as well as $v^{L-1}$ and $h^{L-1}$
    \State Given a two-site Hamiltonian, compute the current energy expectation value (as illustrated in the right panel of Fig.~\ref{fig:PTMRG_scheme})
    \State Using AD and the energy expectation value as a cost function, update the local PEPS tensor to get a lower energy state
    \State Repeat steps $(1)$-$(4)$ until the desired convergence is fulfilled
\end{algorithmic}
In this way, the AD approach allows us to compute the gradient of the energy expectation value as a function of the local PEPS tensor and apply a gradient-based nonlinear optimization method to perform the energy minimization. Based on our experience, we find the quasi-Newton L-BFGS method~\cite{Byrd2006Jul} to be the most efficient and robust approach for this particular case.

 It is worth pointing out a particular pathological problem that can occur when running a gradient-based PEPS optimization\footnote{We have observed that the same issue can occur in AD optimization of infinite PEPS.}. We observe that after several dozen optimization steps, the gradient of the energy sometimes starts to increase dramatically until the line search involved in finding the next optimization step fails. To the best of our knowledge, the origin of the following problem is the discontinuity of the energy function in the PEPS manifold due to truncation errors. Indeed, the energy and associated gradient that is used in the optimization are not the exact PEPS energy. While the latter can be assumed to be smooth, at least for finite systems as is the case here, it cannot be computed exactly. The approximations in evaluating the energy, in this case provided by the truncation step in the PTMRG algorithm, can result in discontinuities (because of e.g. level crossings in the singular value spectrum) and errors which are not variational in origin, \textit{i.e.} which cause the approximated energy to be lower than the exact result. The exact gradient of this approximate energy can guide the optimization process towards exploiting these numerical errors, resulting in unreliable results, such as significantly lower energy with a relatively large gradient norm. The detailed discussion as well as an illustrative example are presented in Appendix~\ref{appendix:optimization_errors}. Therefore, one needs to find a sweet spot for the truncation dimension $\chi$ to balance between sufficient accuracy and manageable computational cost (in both memory and runtime).

\section{Benchmark: transverse field Ising model}

As a first example, we consider the 2-D quantum transverse field Ising model (TFIM) on a square lattice represented by the following Hamiltonian:
\begin{equation}
    \hat{H}_{\text{TFIM}} = - \sum_{\langle ij \rangle} \hat{\sigma}_i^z \hat{\sigma}_j^z - \lambda\sum_{i}\hat{\sigma}_i^x, \quad \lambda > 0,
\end{equation}
where $\sigma^{z},\ \sigma^{x}$ are Pauli matrices, $\langle ij \rangle$ stands for the summation over all nearest-neighbour pairs, and $\lambda$ is the magnitude of the transverse field. For $\lambda=\lambda_{\text{crit}}\approx 3.04438$ there is a phase transition between symmetry-broken and disordered phases in an infinite system. 

We perform PEPS optimization using PTMRG (see Sec. \ref{sec:PTMRG}) and study finite-size scaling for different system sizes $L=3,4,5,..,10$ as well as local PEPS bond dimension $D = 2,3,4$. At the critical point, the ground state energy per site of a finite 2-D quantum system should scale as follows~\cite{PhysRevB.30.322}:
\begin{equation}
    \epsilon_0(L) \approx \epsilon_0 + \frac{\alpha}{L^3} + \dots \;.
    \label{E_crit_eq}
\end{equation}
This behavior is due to the nature of the low-lying energy states, which exhibit a linear dispersion (and thus Lorentz invariance) at the criticality of this model. In contrast, the gapped case leads to exponential corrections to the total energy, or thus, for the energy density~\cite{Klumper1990Mar, Pearce1991Feb, Destri1995Apr, Dugave2015May}:
\begin{equation}
    \epsilon_0(L) \approx \epsilon_0 + \frac{\bar{\alpha} e^{-L/\xi}}{L^2} + \dots .
    \label{E_crit_gapped}
\end{equation}
These relations enable us to extrapolate the ground-state energy density in the thermodynamic limit. 

\subsection{Ground state energy scaling}

\begin{figure}[tb]
    \includegraphics[width=1\columnwidth]{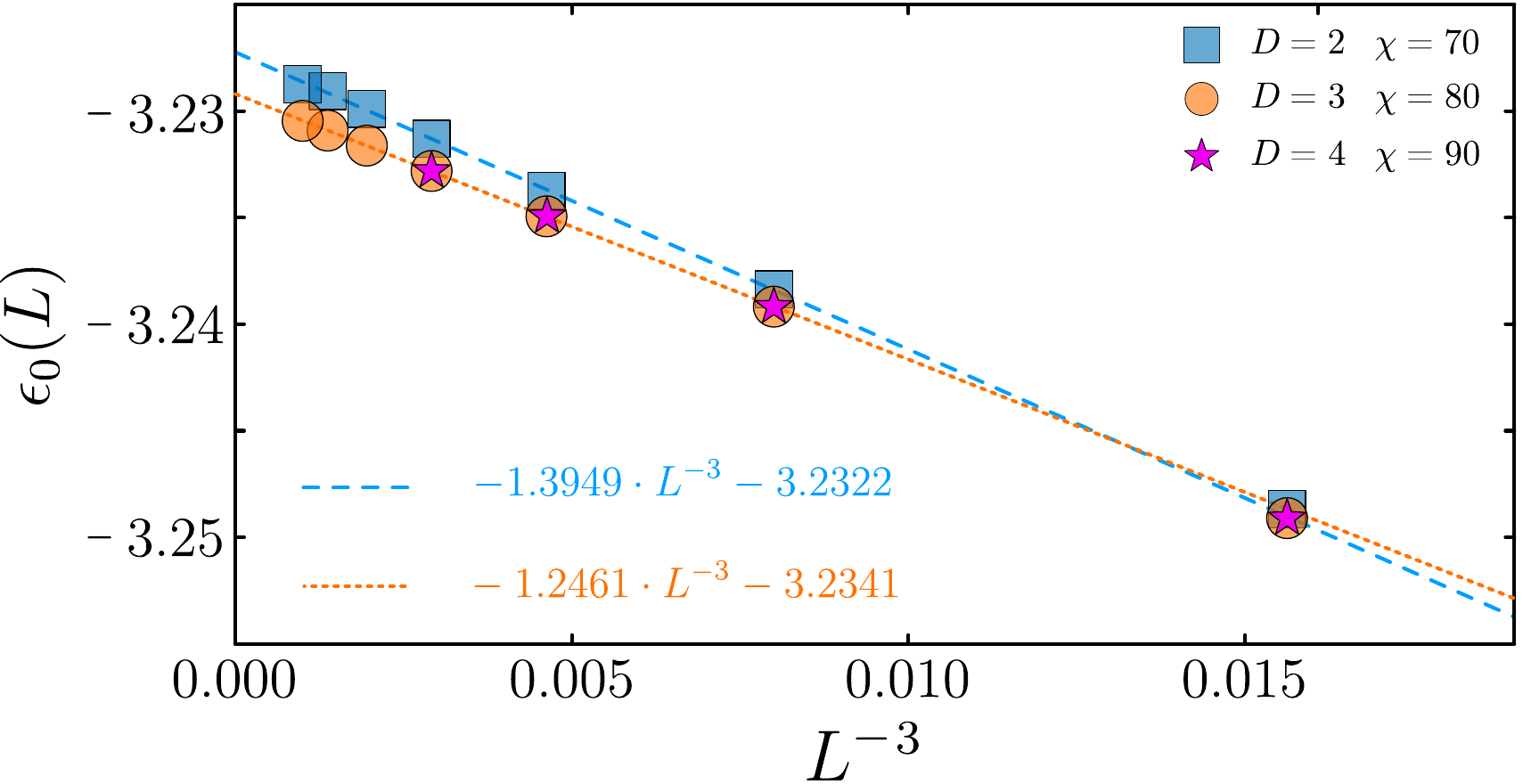}
    \caption{Ground state energy $\epsilon_0$ at the critical point $\lambda_c = 3.04438$ as a function of linear system size $L$ for PEPS with local bond dimension $D=2$ (blue squares), $D=3$ (orange circles) and $D=4$ (magenta stars). Linear fit to $L^{-3}$ scaling shows $\epsilon_0(L \to \infty) \approx -3.2322$ and $-3.2341$ for $D=2,3$, respectively.}
    \label{fig:energy_scaling}
\end{figure}

To verify the scaling behavior stated in Eq.~\eqref{E_crit_eq}, we perform periodic PEPS simulations at $\lambda = \lambda_c = 3.04438$ by varying linear system size $L$. The results are shown in Fig.~\ref{fig:energy_scaling} for $D=2$ and $D=3$ PEPS bond dimensions. To speed up the convergence of our optimization, we start from system size $L=3$; to obtain the ground state of a system with linear size $L$, we use an optimized PEPS tensor for a system with $(L-1) \times (L-1) $ sites as a starting point. Even though starting from a random tensor provides the same result, using results for a smaller system size improves the stability. Furthermore, it decreases the overall computation time of the algorithm until the optimization converges.

Figure~\ref{fig:energy_scaling} shows that $\epsilon_0$ indeed scales as $L^{-3}$ following the CFT argument presented above. A linear fit provides the following ground state energies in the thermodynamic limit:
\begin{equation}
\epsilon_0(L \to \infty) = 
\begin{cases}
    -3.2322, \quad D = 2, \\
    -3.2342, \quad D = 3, 
\end{cases}
\end{equation}
agreeing with infinite PEPS studies presented in~\cite{Rader2018Jul, Vanderstraeten2016Oct}. 


From the ground state energy scaling, we can extract a universal constant for criticality. $\alpha$ in Eq.~\eqref{E_crit_eq} can be decomposed into 
\begin{equation}
\alpha = v\alpha_c,
\label{eq:casimir_scaling}
\end{equation}
where $v$ is a characteristic velocity, and $\alpha_c$ is a (Casimir) constant unique to each universality class. Using the value from a previous study $v \approx 3.323$~\cite{Schuler2016}, we find 
\[
\alpha_c \approx 0.375.
\]
This is consistent with previous studies using exact diagonalization $\alpha_c = 0.35(2)$~\cite{Hamer_2000} and 0.39~\cite{Henkel_1988}.

The detailed list of ground state energies obtained as a function of the linear system size is presented in Table~\ref{tb:energies}. To strengthen our results, we perform the same simulations for local PEPS with bond dimension $D=4$ and different system sizes. Table~\ref{tb:energies} and Fig.~\ref{fig:energy_scaling} (magenta stars) suggest that the ground state energies obtained for $D=4$ are nearly identical to those observed in the $D=3$ case. As the $D=4$ simulations require significantly more time and memory while yielding similar results, in the following, we restrict the virtual bond dimension of PEPS to $D=3$. We find that $D=3$ sufficiently captures the proper scaling. In the last column of Table~\ref{tb:energies}, the results from a snake-pattern finite-size DMRG calculation with $\chi=800$ are shown as a comparison \changes{-- see Appendix~\ref{appendix:DMRG} for a more detailed discussion on the DMRG simulations}.

\begin{table}
\small\addtolength{\tabcolsep}{7pt}
\begin{ruledtabular}
\begin{tabular}{ccccc}
    {$L$} & {$D=2$} & {$D=3$} & {$D=4$} &{DMRG} \\ \midrule
    {$3$}  & {$-3.2837$} & {$-3.2837$} & {$-3.2837$} & {$-3.2837$}\\
    {$4$}  & {$-3.2537$}  & {$-3.2541$} & {$-3.2541$} & {$-3.2541$}\\
    {$5$}  & {$-3.2434$}  & {$-3.2442$} & {$-3.2442$} & {$-3.2442$} \\
    {$6$}  & {$-3.2388$}  & {$-3.2399$} & {$-3.2399$} & {$-3.2399$} \\
    {$7$}  & {$-3.2363$}  & {$-3.2378$} & ${-3.2378}$ & {$-3.2378$}\\
    {$8$}  & {$-3.2349$}  & {$-3.2366$} & &{$-3.2365$} \\
    {$9$}  & {$-3.2341$}  & {$-3.2359$} & &{$-3.2356$}\\
    {$10$} & {$-3.2337$}  & {$-3.2355$} & & {$-3.2347$}\\
    \midrule
    {$L \to \infty$} & {$-3.2322$} & {$-3.2342$} \\
\end{tabular}
\end{ruledtabular}
\caption{Ground state energies of the 2D quantum TFIM at the critical point $\lambda = 3.04438$.}\label{tb:energies}
\end{table}

\subsection{Magnetization for finite \texorpdfstring{$h_z$}{a}}

Once we have obtained optimized PEPS tensors representing the ground state of the 2D TFIM, it is quite natural to extend our approach and calculate the magnetization per site $m_z=\braket{\sigma^z}$ along the $z$-axis. To proceed, we follow the same approach to evaluate the two-site expectation value as we did in PTMRG, only replacing the two-site Hamiltonian with a single-site $\sigma^z$ operator. 

Since finite-size systems do not support spontaneous symmetry breaking, $m_z$ just trivially equals zero for all $\lambda$. Therefore, to break the $\mathbb{Z}_2$ symmetry explicitly in TFIM, we introduce a small longitudinal $h_z$ field. As a result, the total Hamiltonian takes the following form:
\begin{equation}
\hat{H} = \hat{H}_{\text{TFIM}} - h_z \sum_i \hat{\sigma}_i^z, \quad h_z > 0.
\end{equation}

\begin{figure}[tb]
    \includegraphics[width=\columnwidth]{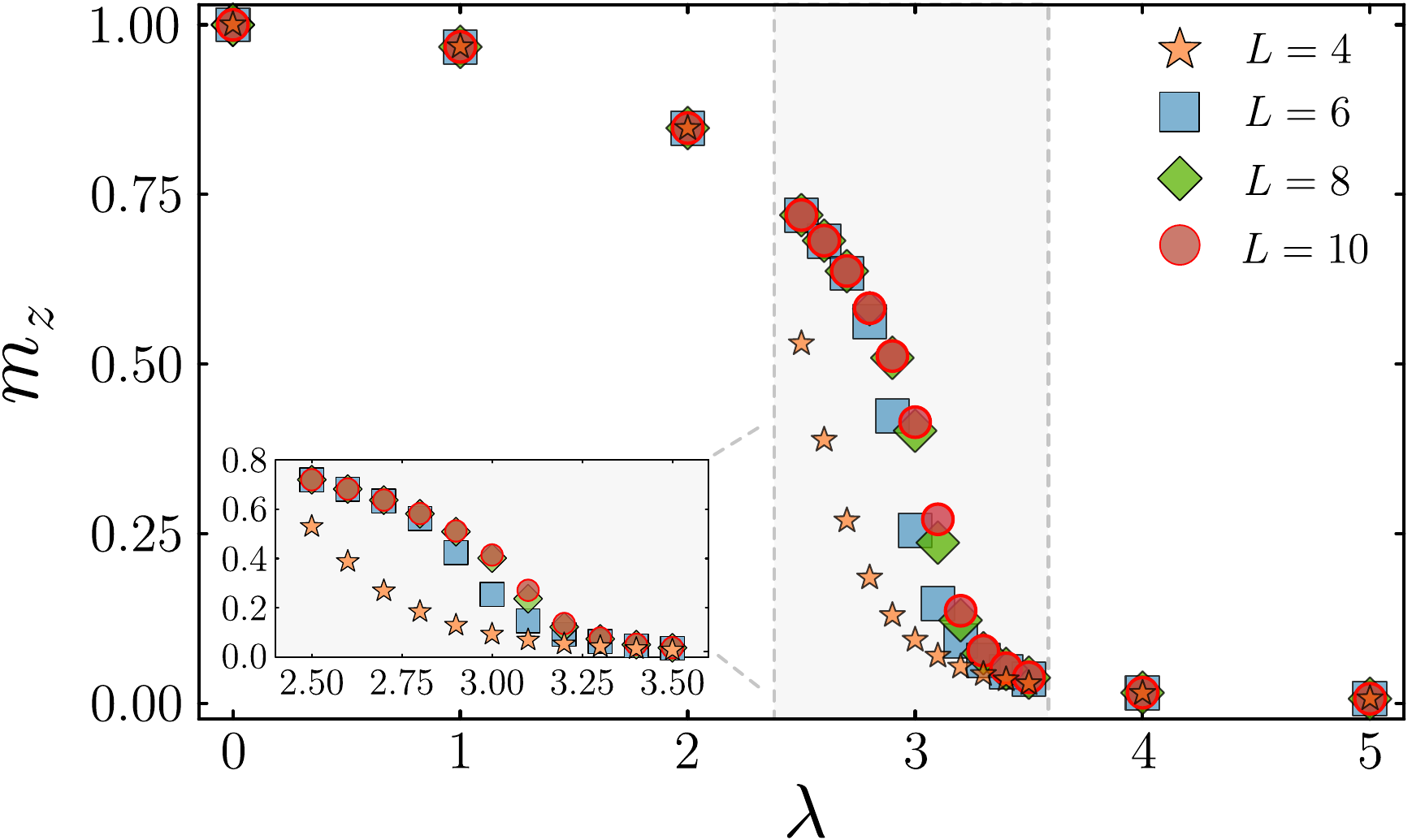}
    \caption{The magnetization under a small uniform magnetic field with $h_z = 10^{-2}$ various system sizes $L = 4,6,8,10$. }
    \label{fig:magn_over_lambda}
\end{figure}

By setting $h_z=10^{-2}$ and PEPS bond dimension $D=2$, we perform PTMRG minimization and, using optimized PEPS tensors, calculate $m_z$ as a function of $\lambda$ and compare different linear system sizes as shown in Fig.~\ref{fig:magn_over_lambda}. As expected, for $\lambda = 0$, we obtain 
a symmetry-broken state when all the spins are aligned along the $z$-axis and magnetization per site $m_z=1$. By increasing $\lambda$, $m_z$ goes to zero, and the critical point appears around $\lambda \approx 3.0$.

\section{Extracting scaling dimension with finite-size effects}
Using our periodic PEPS algorithm, we can now start to extract the critical properties for a given quantum critical point. In general, we are interested in the scaling dimensions of the conformal field theory describing the low-energy physics around the critical point. These scaling dimensions appear in the correlation functions as
\begin{equation}
    \langle \hat{\Phi}_g(0) \hat{\Phi}_g(r)\rangle \propto \frac{1}{|r|^{2\Delta_g}},\label{eq:scaling_correlation}
\end{equation}
where the scaling dimension is denoted as $\Delta_g$. This comes from the transformation rules of the operator $\hat{\Phi}_g(r) \rightarrow b^{-\Delta_g}\hat{\Phi}_g(r')$ under scale transformations $r= br'$. Equation~\eqref{eq:scaling_correlation} is obtained when we set $b=r$ so that $\langle \hat{\Phi}_g(0) \hat{\Phi}_g(r)\rangle \propto r^{-2\Delta_g}\langle\hat{\Phi}_g(0) \hat{\Phi}_g(1)\rangle$.

Here we will extract these scaling dimensions by perturbing the critical Hamiltonian with an operator $\hat{\Phi}$, and reading off the scaling dimension from the scale dependence of the amplitude of the perturbation. In effective field theory, this can be expressed as 
\begin{equation}
    \hat{H}_{\text{pert}} = \hat{H}^*_{\text{CFT}} + \sum\limits_{n} g_n\int\limits_{0}^{Lx,Ly}\hat{\Phi}_n({\mathbf{r}})d^2 \mathbf{r},
\end{equation}
where $n$ represents all possible perturbations with corresponding $\hat{\Phi}_n({\mathbf{r}})$ operators, and $g_n$ is the running couplings. In the case of the 2D quantum TFIM, there are two relevant perturbations, $\hat{\Phi}_\epsilon({\mathbf{r}})$ and $\hat{\Phi}_\sigma({\mathbf{r}})$ that respectively correspond to uniform transverse ($\hat{\sigma}^x$) and longitudinal ($\hat{\sigma}^z$) perturbations from the critical coupling $\lambda_c$ and 0. In the following, we neglect the contributions from the irrelevant operators as they are significantly smaller than those of relevant operators.

\begin{figure}[tb]
    \includegraphics[width=1\linewidth]{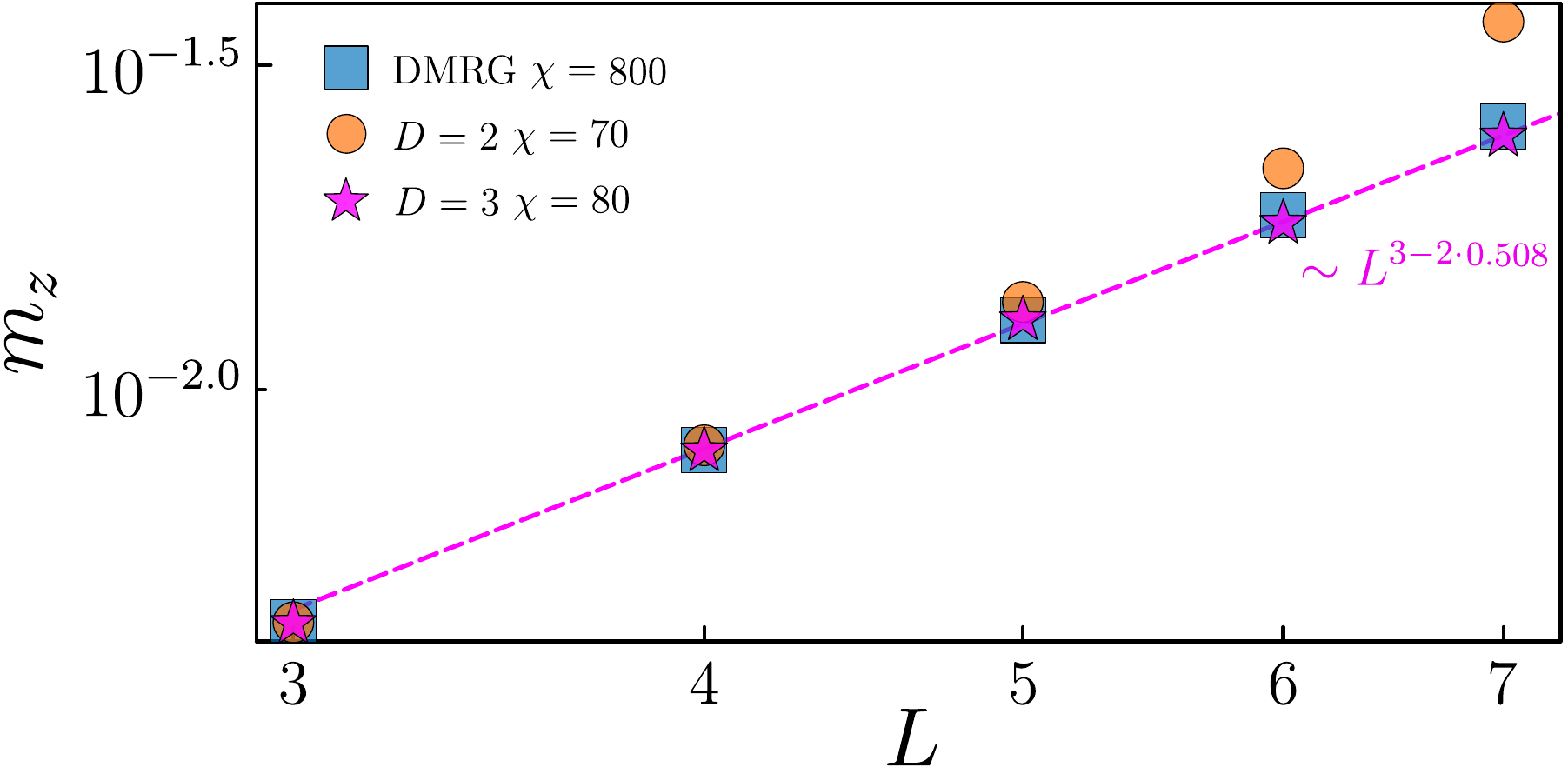}
    \caption{Magnetization per site $m_z$ as a function of system size $L$ and longitudinal perturbation $h_z = 10^{-3}$ for PEPS bond dimension $D=2$ (orange circles) and $D=3$ (magenta stars) compared with DMRG simulation (blue squares). $\Delta_\sigma = 0.508$ is extrapolated from a polynomial fit (magenta dashed line).}
    \label{fig:mz_scaling}
\end{figure}

Let us consider the $\hat{\Phi}_\sigma({\mathbf{r}})$ perturbation. In this case, the Hamiltonian takes the following form:
\begin{align*}
    \hat{H}_{\text{pert}}^{\sigma} &= \hat{H}^*_{\text{CFT}} + g_\sigma\hat{V}\\
    &=  \hat{H}^*_{\text{CFT}} + g_\sigma \int\limits_{0}^{Lx,Ly}\hat{\Phi}_\sigma({\mathbf{r}})d^2 \mathbf{r}.
\end{align*}
Let $\ket{\psi^0}, \ket{\psi^1}, \ket{\psi^2},...$ and $E_0, E_1,E_2,...$ be eigenstates and eigenvalues in ascending order of nonperturbed Hamiltonian $\hat{H}^*_{\text{CFT}}$. Then, the modified ground state of the perturbed $\hat{H}_{\text{pert}}^{\sigma}$ according to standard perturbation theory is
\begin{equation}
    \ket{\psi^{0}}_\sigma = \ket{\psi^0} + g_\sigma\frac{\bra{\psi^{1}}\hat{V}\ket{\psi^0}}{E_0 - E_1}\ket{\psi^1} + ...,
\end{equation}
where we keep only linear order terms $O(g_\sigma)$. Here, the ground state $\ket{\psi^0}$ and the first excited state $\ket{\psi^1}$ are even and odd under the spin-flip symmetry. Given that the longitudinal magnetic field $\hat{V}$ is odd under the same symmetry, it allows for nonzero magnetization from the cross terms of these two states. Using this, the magnetization per site $m_z$ becomes
\begin{align}
    m_z &= \bra{\psi^0_\sigma}\hat{\sigma}^z\ket{\psi^0_\sigma} \nonumber\\
    &\simeq \langle\psi^0|\hat{\sigma}^z|\psi^1\rangle g_\sigma\left(\frac{\bra{\psi^{1}}\hat{V}\ket{\psi^0}}{E_0 - E_1}\right)+H.c.\nonumber\\
    &\propto \mathrm{Re}\left(2\langle\psi^0|\hat{\sigma}^z|\psi^1\rangle g_\sigma\frac{\bra{\psi^{1}}\hat{V}\ket{\psi^0}}{E_0 - E_1}\right).
\end{align}

The first term $\langle\psi^0|\hat{\sigma}^z|\psi^1\rangle$ scales as $L^{-\Delta_\sigma}$, following the scaling of the $\Psi_\sigma$ operator. Similarly, the matrix element $\bra{\psi^{1}}\hat{V}\ket{\psi^0}$ can be evaluated with the scaling relation as $\bra{\psi^{1}}\int_0^L d^2r\hat{\Phi}_\sigma \ket{\psi^0} \propto L^{2-\Delta_\sigma}$ and $E_0 - E_1 \propto L^{-1}$ at the critical point. As a result, we get the following scaling for the magnetization per site as
\begin{equation}
    m_z \propto L^{3 - 2\Delta_\sigma}.\label{eq:mz_scaling}
\end{equation}
In more generic cases, the exponents consist of the scaling of the operator and the amplitude of the perturbation/running couplings. The first one scales as $L^{-\Delta_\sigma}$, while the second $L^{d-\Delta_\sigma}$ for the systems with $d$ space-time dimensions. The latter is nothing but the RG dimension of the operator $\hat{\Phi}_\sigma$. 
Similarly, we can also read out the RG dimensions of other operators by perturbing the critical Hamiltonian with the corresponding operators.

To confirm the scaling behavior in Eq.~\eqref{eq:mz_scaling}, we perform PTMRG optimization of periodic PEPS with bond dimensions $D=2,3$ at the critical point $\lambda=3.04438$ and longitudinal perturbations $h_z = 10^{-3}$. We calculate the magnetization $m_z$ per site for different linear system sizes $L$, as shown in Fig.~\ref{fig:mz_scaling}. We confirm that we find a power-law behavior as a function of $L$ and $m_z$ for our PEPS simulations with bond dimension $D = 3$. Using Eq.~\eqref{eq:mz_scaling}, one can read off the scaling dimension from its power-law exponent as
\begin{equation}
\Delta_\sigma = 0.508,
\end{equation}
in a reasonable agreement with the value in the literature \changes{$\Delta_\sigma = 0.518148806(24)$ and $0.524$ respectively obtained from conformal bootstrap~\cite{chang2024bootstrapping} and fuzzy sphere regularization~\cite{fuzzy_sphere}}. 

\begin{figure}[tb]
    \includegraphics[width=1\columnwidth]{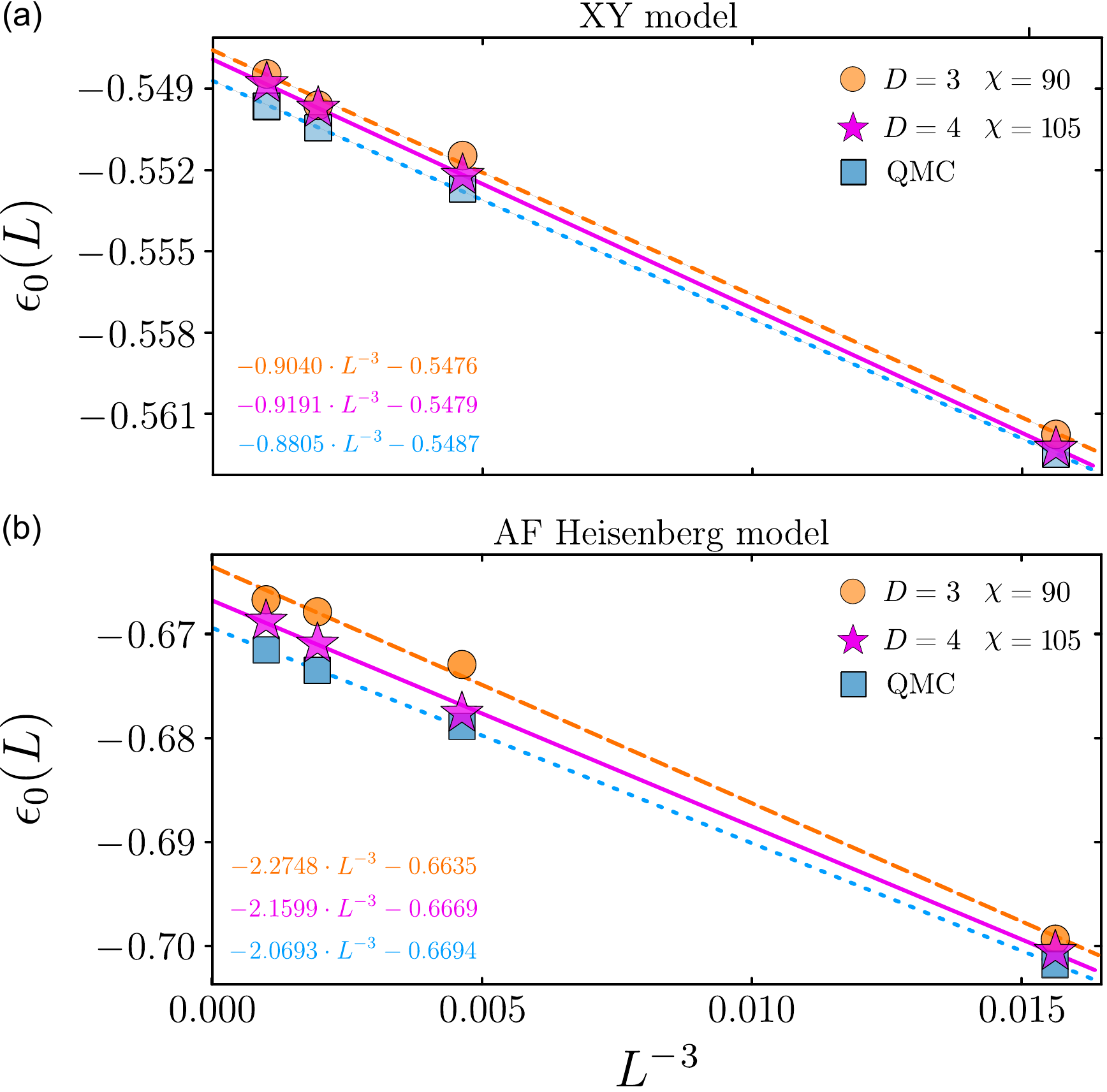}
    \caption{Ground state energy $\epsilon_0$ of 2D (a) XY  and (b) AF Heisenberg models as a function of linear system size $L$ for $\mathbb{Z}_2$-PEPS with local bond dimension $D=3$ (orange circles) and $D=4$ (magenta stars). Blue squared points show QMC results presented in (a)~\cite{Sandvik1999Sep} and (b)~\cite{Sandvik1997Nov}. Linear fit to $L^{-3}$ scaling shows (a) $\epsilon_0(L \to \infty) \approx -0.5476$ and $-0.5479$ and (b) $\epsilon_0(L \to \infty) \approx -0.6635$ and $-0.6669$ for $D=3,4$, respectively.}
    \label{fig:xy_and_heisen_scaling}
\end{figure}

\section{XY and Heisenberg models}

In this section, we consider the generalized quantum Heisenberg model on a 2D square lattice, which can be represented as:
\begin{equation}
    \hat{H} = \sum_{\langle ij\rangle}S^x_i  S^x_j + S^y_iS^y_j + \Delta S^z_i S^z_j,
\end{equation}
where $\{S^{x,y,z}\}$ are the spin-1/2 operators. $\Delta = 0$ correspond to the XY ($O(2)$) model, while $\Delta = 1$ the antiferromagnetic Heisenberg ($O(3)$) model. In the thermodynamic limit, these models spontaneously break their continuous symmetry, creating one (two) Nambu-Goldstone bosons as gapless excitations for XY (AF Heisenberg), respectively. Since it is known that these systems are critical, we study the scaling behavior of the ground state energy $\epsilon_0$ as a function of a linear system size $L$.

\begin{table}
\centering
\vspace{10pt}
\begin{ruledtabular}
\begin{tabular}{c|c|c|c|c|c|c} 
\multicolumn{1}{l}{} & \multicolumn{3}{c|}{XY model}  & \multicolumn{3}{c}{AF Heisenberg model} \\ 
\midrule
$L$ & $D = 3$ & $D = 4$ & QMC & $D = 3$ & $D = 4$ & QMC \\ 
\midrule
$4$ & {$-0.5618$} & {$-0.5623$} & {$-0.5625$} & $-0.6993$ & $-0.7005$ & $-0.7018$  \\
$6$ & {$-0.5515$}  & {$-0.5522$} & {$-0.5527$} & $-0.6729$ & $-0.6777$ &  $-0.6789$ \\
$8$ & {$-0.5496$}  & {$-0.5498$} & {$-0.5504$}& $-0.6679$ & $-0.6710$ &  $-0.6735$ \\
$10$  & {$-0.5485$}  & {$-0.5489$} & {$-0.5496$} & $-0.6667$ & $-0.6688$ & $-0.6715$ \\
\midrule
$\to \infty$ & {$-0.5476$} & {$-0.5479$} & {$-0.5487$}& $-0.6635$ & $-0.6669$ &  $-0.6694$ \\
\end{tabular}
\end{ruledtabular}
\caption{Ground state energies of the 2D quantum XY and AF Heisenberg models obtained by PTMRG optimization and compared to QMC results as a function of a linear system size $L$.}\label{tb:XY_and_Heisen}
\end{table}

To proceed, we perform PTMRG optimization of the ground state energy for PEPS bond dimensions $D = 2,3,4$ and impose $\mathbb{Z}_2$-symmetry. To work with single-site unit cells, we perform a sublattice rotation as well. We observe that $D = 2$ PEPS ansatz is insufficient to capture the scaling behavior since the final optimized ground state energy remains constant as the linear system size $L$ increases. Meanwhile, for $D=3,4$ the obtained ground state energies are shown in Table \ref{tb:XY_and_Heisen} together with QMC results presented in Refs.~\cite{Sandvik1999Sep, Sandvik1997Nov} and extrapolation to the thermodynamic limit $L \to \infty$. 

Figure~\ref{fig:xy_and_heisen_scaling} shows the optimized ground state energy as a function of the system size $L$ both for the (a) XY and (b) AF Heisenberg models. As expected from 3-D CFT, the energy density scales as $L^{-3}$ and approaches the QMC results with increasing PEPS bond dimension $D$.
The Casimir constants for the XY and AF Heisenberg models are found to be,
\begin{equation}
\begin{cases}
        \ \alpha_c^{XY} \approx 0.813 \quad \text{for XY model}, \\
        \ \alpha_c^{HB} \approx 1.302 \quad \text{for AF Heisenberg model},
\end{cases}
\end{equation}
after substituting the corresponding spin-wave velocities $v_{XY} = 1.123$ and $v_{HB} = 1.658 47$ in the literatures~\cite{Sandvik1999Sep,Sen2015Nov}. Our estimations are in good agreement with $\alpha_{QMC}^{XY} = 0.781$  and $\alpha_{QMC}^{HB} = 1.246$ obtained from QMC simulations in~\cite{Sandvik1999Sep,Sandvik1997Nov}. The list of the calculated Casimir constants is presented in Table~\ref{tb:casimir_factors} together with the values from previous studies. For 2-D quantum critical systems on the torus, those values can be used to detect the universality class in future studies.

\begin{table}
\centering
\vspace{10pt}
\begin{ruledtabular}
\begin{tabular}{c|c|c} 
\multicolumn{3}{c}{Casimir critical constant $\alpha_c$} \\
\hline
 Model & PTMRG & previous works \\ 
\hline
Transverse-field Ising  & {$0.375$} & {$0.35$}~\cite{Hamer_2000}\\ 
\hline
XY  & ${0.813}$  & {$0.781$}~\cite{Sandvik1999Sep} \\
\hline
AF Heisenberg  & ${1.302}$  & {$1.246$}~\cite{Sandvik1997Nov} \\
\end{tabular}
\end{ruledtabular}
\caption{Casimir critical factor $\alpha_c$ extracted from the ground energy scaling with linear system size for 2D QTIM, XY, and AF Heisenberg models.}\label{tb:casimir_factors}
\end{table}

\section{Conclusion and outlook}

In summary, we propose a novel PTMRG algorithm to contract 2-D tensor networks with periodic boundary conditions at a significantly lower computational cost compared to the well-established TRG and HOTRG schemes. By acquiring data for linearly growing system sizes, we demonstrate the performance of PTMRG for problems in 2-D statistical mechanics and benchmark the universality of the transfer matrix spectrum of the 2D classical Ising model at criticality.  

For 2-D quantum lattice models, we perform the variational optimization using AD-computed gradients on top of the PTMRG contraction scheme, thus obtaining a fully periodic and uniform PEPS algorithm. We demonstrate finite-size scaling of ground state energies for the 2-D transverse-field Ising model on a torus both for gapless and gapped phases. We extend the algorithm towards the extraction of scaling dimensions via finite-size effects. The torus geometry as a PEPS tensor network enables us to support our numerics with effective field theory arguments. Finally, we perform finite-size scaling of the ground state energies for XY and Heisenberg models, which are quite challenging for PEPS due to long-range order and quantum fluctuations, and demonstrate the performance of our approach as compared to state-of-the-art QMC results. \changes{By further improving the gradient computation using, e.g., a more performant AD engine and checkpointing strategies, we will likely be able to further push the quality of these results.}

We believe that our findings can support ongoing research in the tensor network simulation of  quantum criticality in 2+1 dimensions. In particular, the periodic finite-size approach is complementary to finite-entanglement scaling for infinite PEPS. The next step would consist of going beyond ground states and computing excitation spectra on the torus with PEPS, as torus spectroscopy can be a very powerful method for revealing the scaling dimensions of exotic phase transitions~\cite{Schuler2016, Whitsitt2017, Schuler2021, Wietek2024}. Generalizing the excitation ansatz to the setting of periodic PEPS would be very natural, as this approach was shown to work well for infinite PEPS~\cite{Vanderstraeten2015, Vanderstraeten2019, Ponsioen2020} as well as periodic MPS \cite{Pirvu2012,Zou2018, Zou2020, Zou2020b}.

Finally, it would be very interesting to explore boundary conditions different from the simple periodic ones we have considered here. Our scheme should be able to adopt the twisted boundary conditions phrased in terms of tube algebras and tensor networks~\cite{BULTINCK2017183,PRXQuantum.5.010338}, which is important to probe all sectors of the theory.

\section*{Acknowledgement}

The authors thank Andreas L\"auchli for inspiring discussions. G.F. was supported by the grant BOF23/GOA/021 from Ghent University. A.U. was supported by the MERIT-WINGS Program at the University of Tokyo, the JSPS fellowship (DC1), BOF-GOA (grant No. BOF23/GOA/021), and Watanabe Foundation. This work was supported by the Research Foundation Flanders (FWO) via grant GOE1520N,  EOS (grant No.
40007526), IBOF (grant No. IBOF23/064), and BOF-
GOA (grant No. BOF23/GOA/021). All the simulations were performed using the TensorKit.jl package~\cite{tensorkit}.

\changes{\textbf{Note added:} After completion of this work, we became aware of Ref.~\onlinecite{Lan_2019}, which presents a tensor renormalization scheme that is very similar to the one presented in this work, also consisting of a corner and horizontal and vertical edge tensors and a linear rather than exponential coarse-graining strategy. The specific update rules for the tensors are different, and this work further factorizes the four-leg corner tensor into two three-leg tensors, which further reduces the computational complexity to $O(\chi^4)$. This scheme was only tested on the partition function of the classical Ising model, so it would be useful to compare both schemes in the context of PEPS optimization.}

\begin{appendix}
\section{2D Quantum Transverse field Ising model: gapped phase}

\begin{figure}[tb]
    \includegraphics[width=1\columnwidth]{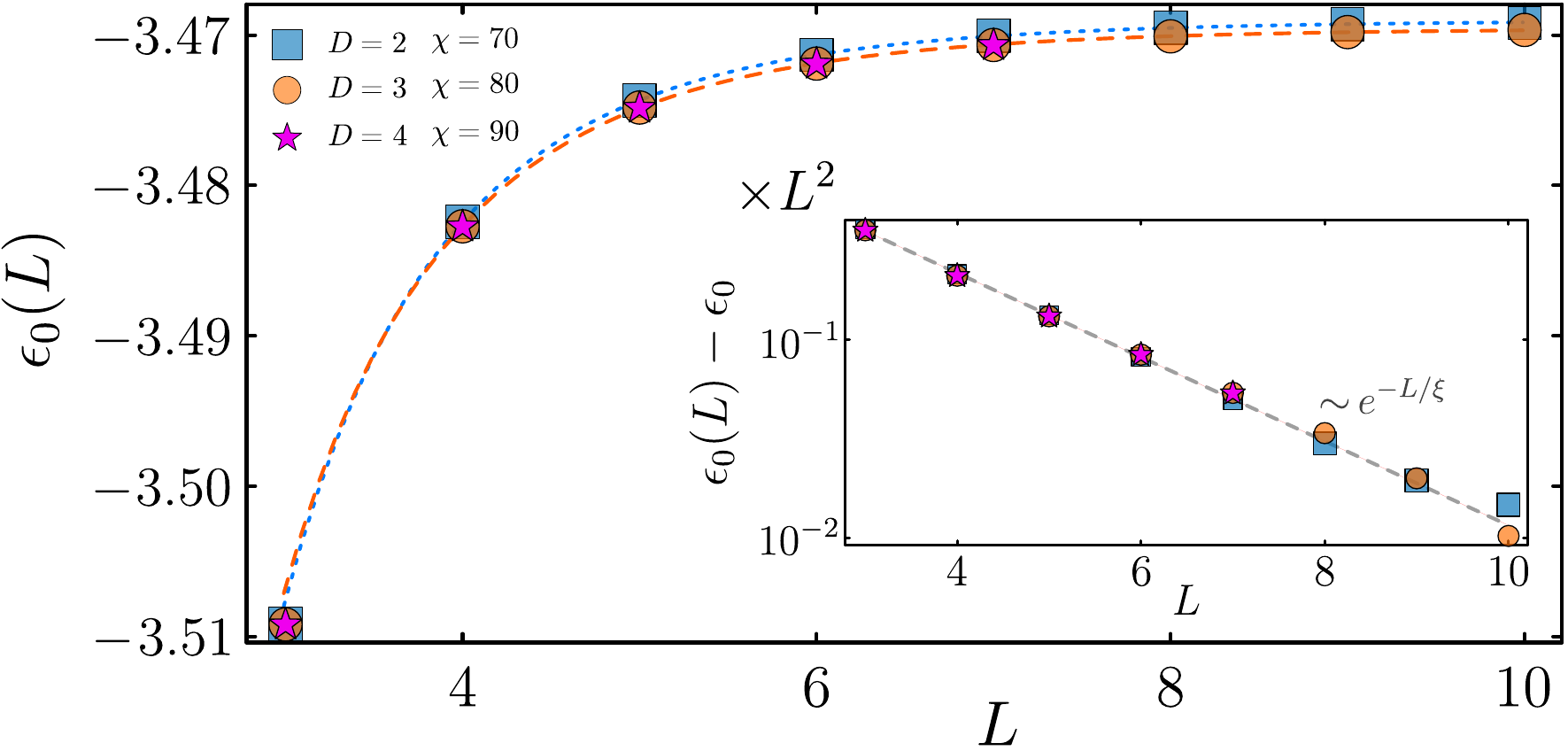}
\caption{Ground state energy $\epsilon_0$ at the gapped phase $\lambda = 3.3$ as a function of linear system size $L$ for PEPS with local bond dimension $D=2$ (blue squares), $D=3$ (orange circles), $D=4$ (magenta stars). (Insert) Linear fit of the difference between extrapolated value $\epsilon_0$ when $L\to \infty$ and $\epsilon_0(L)$ scales exponentially with the system size $L$.}
    \label{fig:energy_scaling_gapped}
\end{figure}

The ground state energy of gapped phases scales differently from that of critical systems as in Eq.~\eqref{E_crit_gapped} in the main text. To confirm this exponential decay of the finite-size effects, we simulate the transverse field Ising model away from the critical point $\lambda = 3.3$. The obtained ground state energy indeed converges quickly with increasing the system size, as shown in Fig.~\ref{fig:energy_scaling_gapped}. As for more precise scaling, we extrapolate the ground state energy in the thermodynamic limit from the finite-size data according to Eq.~\eqref{E_crit_gapped}. The result is illustrated in the insert, which confirms that the finite size correction to the ground state is exponential as $\epsilon_0(L)-\epsilon_0(\infty)~\sim \frac{e^{-L/\xi}}{L^2}$. In addition, we list the numerical data with respect to the bond dimension $D$ in Table~\ref{tb:energies_gapped}. The energy is almost identical for all $D$, indicating that the gapped phases do not require a high bond dimension.

\begin{table}
\vspace{10pt}
\begin{ruledtabular}
\begin{tabular}{cccc} 
    {$L$} & {$D=2$} & {$D=3$} & {$D=4$} \\ \midrule
    {$3$}  & {$-3.5092$}  & {$-3.5092$} & {$-3.5092$} \\
    {$4$}  & {$-3.4824$}  & {$-3.4827$} & {$-3.4827$} \\
    {$5$}  & {$-3.4743$}  & {$-3.4748$} & {$-3.4748$} \\
    {$6$}  & {$-3.4713$}  & {$-3.4719$} & {$-3.4719$} \\
    {$7$}  & {$-3.4700$}  & {$-3.4707$} & {$-3.4707$}\\
    {$8$}  & {$-3.4695$}  & {$-3.4701$} \\
    {$9$}  & {$-3.4693$}  & {$-3.4698$}\\
    {$10$} & {$-3.4692$}  & {$-3.4697$}\\
    \midrule
    {$L \to \infty$} & {$-3.4690$} & {$-3.4696$} \\
\end{tabular}
\end{ruledtabular}
\caption{Ground state energies of the 2D quantum TFIM in the paramagnetic phase at $\lambda = 3.3$.}\label{tb:energies_gapped}
\end{table}

\section{Discontinuities during gradient optimization}\label{appendix:optimization_errors}

\begin{figure}[tb]
    \vspace{10pt}
    \includegraphics[width=1\columnwidth]{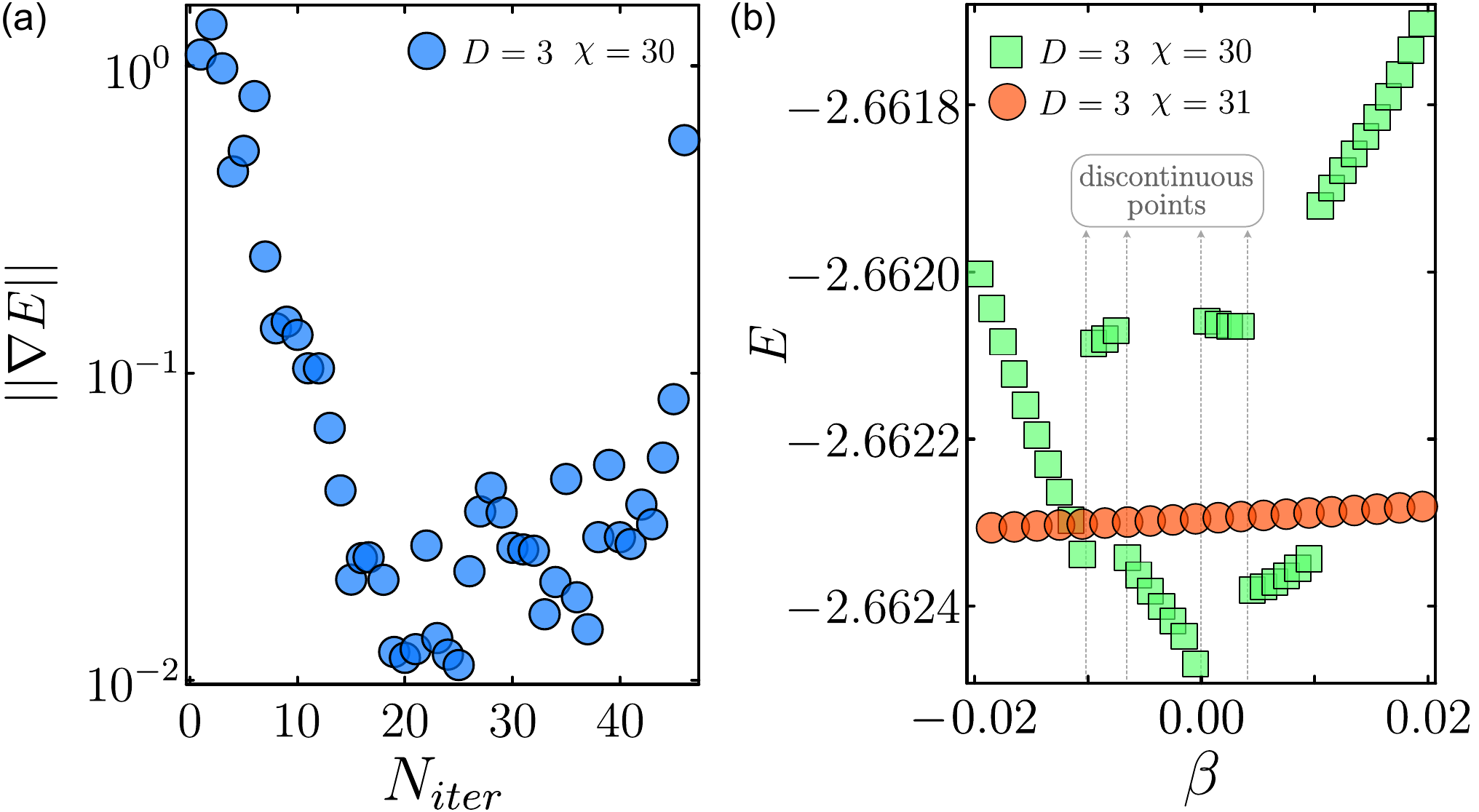}
    \caption{An example of the discontinuity appearing in the (approximate) energy when 
 we use an insufficient truncation dimension $\chi$. (a) The gradient of the AF Heisenberg ground state energy as a function iteration number $N_{iter}$. The gradient exhibits an abrupt increase after 50 iterations. (b) The energy landscape at this unfavorable point. $\beta$ is a one-dimensional parameter on the parameter space of the PEPS tensor. For $\chi=30$, the energy is discontinuous, preventing the possibility of using gradient optimization, as the linesearch will fail. This issue is fixed once you have a sufficiently large truncation dimension.}
    \label{fig:disc_points}
\end{figure}

In the main text, we highlight the pathological issue that could appear while performing gradient-based optimization with PEPS algorithms that involve truncation. This is due to the possible discontinuity of the approximate energy evaluation as a function of the variational parameters in the PEPS tensors when we employ insufficient truncation dimension, which can ultimately be traced down to level crossings in the singular value spectra on which the truncation will act. We observe that this issue arises, in particular, when studying more challenging models with larger correlation lengths.
To demonstrate it, we simulate the AF Heisenberg model with the PEPS bond dimension $D=3$ and truncation dimension $\chi = 30$. Here, the bond dimension of the double-layer PEPS tensor is $D^2=9$, which induces enormous truncation errors during the contraction of the tensors. 

Figure~\ref{fig:disc_points}(a) shows the obtained gradient of the energy $\nabla E$ as a function of optimization steps $N_{iter}$. We observe that the gradient decreases up to $\approx 25$ iteration steps, but then the gradient starts increasing until the gradient optimization breaks down and can not find the direction to optimize the energy, while the gradient norm is still relatively large, taking values around $10^{-1}$. To see what is causing this failure, we compute the energy landscape for this particular PEPS tensor that breaks the line search, which amounts to a one-dimensional parameter search $\beta$ to find the energy minimum. Figure~\ref{fig:disc_points}(b) demonstrates how the energy behaves along the gradient direction $\beta \cdot \nabla E_A$ where $\beta \in [-0.02,0.02]$. For $\chi = 30$, there are several discontinuous points where the energy $E$ experiences sudden jumps, as denoted with gray vertical lines. These discontinuities are not representative of those of the exact energy function, which can be assumed to be a smooth function of the variational parameters, but can unfortunately not be computed efficiently. Hence, these discontinuities result from the approximation, more specifically, the truncation step, which is here based on a singular value decomposition. A level crossing of singular values around the truncation dimension will cause a sudden jump in the associated singular vectors that are kept versus discarded. These discontinuities will confuse the line search process, causing it to fail. Furthermore, with AD we obtain the exact gradient of this approximate energy landscape, and the gradient optimization can thus be guided towards directions with larger truncation errors, if these errors have the effect of lowering the energy approximation (to nonphysical values). As illustrated in Fig.~\ref{fig:disc_points}(b), increasing the bond dimension to $\chi = 31$ resolves the discontinuities in this region of the energy landscape, but might of course cause discontinuities elsewhere. Hence, the bond dimension $\chi$ needs to be sufficiently high to avoid discontinuities altogether, or at least make them very small compared to the natural steps taken in the optimization process. Hence, a balance needs to be found between sufficient accuracy and manageable computational cost, as increasing $\chi$, of course, affects the overall computation time and memory required, in particular in combination with the use of AD for computing the gradient via back-propagation.

\changes{
\section{Computational cost of a single PTMRG step}\label{appendix:cost_PTMRG}
\begin{figure}[tb]
    \includegraphics[width=1\columnwidth]{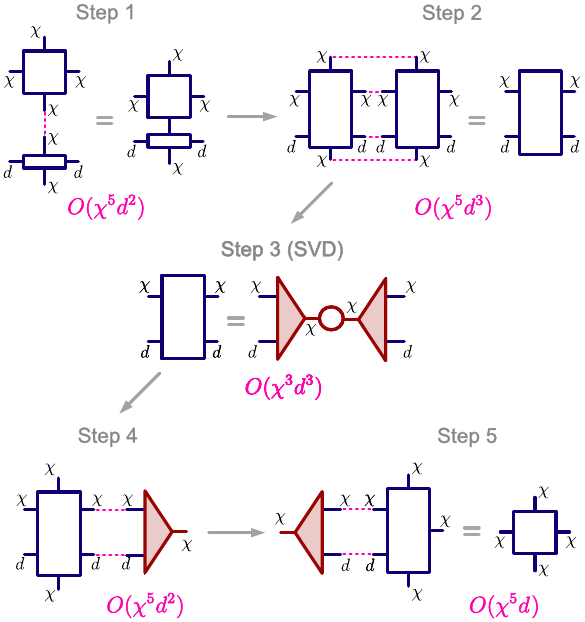}
\caption{An example of a single PTMRG step that updates corner tensor with computational cost per each contraction sub-step.}
    \label{fig:cost_per_pTMRG_step}
\end{figure}
One of the main findings of the manuscript is the reduction of the computational cost of a single RG step. Namely, by linearly increasing the system size, we can achieve $O(\chi^5)$ scaling instead of $O(\chi^6)$ and $O(\chi^7)$ for TRG and HOTRG, respectively. In this section, we confirm this statement by considering a single RG step to update the corner tensor as explained in the main text.}

\changes{Let us consider a corner tensor with four legs (each with a bond dimension $\chi$) and a horizontal edge tensor (top/bottom legs with bond dimension $\chi$ and left/right legs with bond dimension $d$). The steps within a single PTMRG step are shown in Fig.~\ref{fig:cost_per_pTMRG_step} and explained below:
\begin{algorithmic}[1]
    \State Contract the corner and edge tensor along their shared leg of dimension $\chi$ → cost $O(\chi^5d^2)$
    \State Contract the resulting tensor with its adjoint along 3 legs of dimension $\chi$ and one leg of dimension $d$ → cost $O(\chi^5d^3)$
    \State Perform an SVD on the resulting $(\chi d) \cdot (\chi d)$ matrix → cost $O(\chi^3d^3)$
    \State Use the resulting isometry to absorb the edge tensor into the corner, by first contracting it with the tensor from Step 1 along a leg of dimension $\chi$ and one leg of dimension $d$ → cost $O(\chi^5d^2)$
    \State Now also contract with the isometry from the other side to obtain the updated corner tensor → cost $O(\chi^5d)$
\end{algorithmic}
As a result, the leading order of the computational cost with respect to the truncation dimension $\chi$ is $O(\chi^5)$. The fact that the former works with tensors that have the bond dimension $d$ of the initial local $4$-leg tensors is an intuitive explanation for why the numerical cost of PTMRG is $\chi^2$ cheaper than that of HOTRG. Updates for edge tensors require even less cost since by definition the tensors have two legs with bond dimension $\chi$ and two legs with bond dimension $d$. For the non-symmetric case, we can generalize this strategy by using an “oblique projector" as described in Refs.~\onlinecite{Yang2017,Wang2011ClusterUF,PhysRevB.100.035449,Corboz2014}.}

\changes{
\section{DMRG simulation}\label{appendix:DMRG}
\begin{table*}
\small\addtolength{\tabcolsep}{2pt}
\begin{ruledtabular}
\begin{tabular}{cccccccc}
    {$L$} & {$\chi=100$} & {$\chi=150$} & {$\chi=200$} & {$\chi=250$} & {$\chi=300$} & {$\chi=500$} & {$\chi=800$}\\ \midrule
    {$3$}  & {$-3.2837$} & {$-3.2837$} & {$-3.2837$} & {$-3.2837$} & {$-3.2837$} & {$-3.2837$} & {$-3.2837$}\\
    {$4$}  & {$-3.2541$} & {$-3.2541$} & {$-3.2541$} & {$-3.2541$} & {$-3.2541$} & {$-3.2541$} & {$-3.2541$}\\
    {$5$}  & {$-3.2441$} & {$-3.2442$} & {$-3.2442$} & {$-3.2442$} & {$-3.2442$} & {$-3.2442$} & {$-3.2442$}\\
    {$6$}  & {$-3.2396$} & {$-3.2398$} & {$-3.2399$} & {$-3.2399$} & {$-3.2399$} & {$-3.2399$} & {$-3.2399$}\\
    {$7$}  & {$-3.2368$} & {$-3.2373$} & {$-3.2375$} & {$-3.2376$} & {$-3.2377$} & {$-3.2378$} & {$-3.2378$}\\
    {$8$}  & {$-3.2345$} & {$-3.2354$} & {$-3.2359$} & {$-3.2361$} & {$-3.2363$} & {$-3.2365$} & {$-3.2365$}\\
    {$9$}  & {$-3.2323$} & {$-3.2338$} & {$-3.2344$} & {$-3.2348$} & {$-3.2351$} & {$-3.2356$} & {$-3.2356$}\\
    {$10$}  & {$-3.2305$} & {$-3.2321$} & {$-3.2330$} & {$-3.2336$} & {$-3.2340$} & {$-3.2347$} & {$-3.2347$}\\
\end{tabular}
\end{ruledtabular}
\caption{Ground state energies of the 2D quantum TFIM at the critical point $\lambda = 3.04438$ obtained via periodic DMRG, $\chi$ states for the truncation bond dimension.}\label{tb:DMRG_data}
\end{table*}}
\changes{In the main text, we present the ground state energies of the 2-D quantum transverse field Ising model (TFIM) on a square lattice at the critical point to illustrate the performance of the proposed PEPS optimisation algorithm. As no exact results are known, we also provide DMRG results for the same system. Here, we snake the MPS (with open boundary conditions) as in conventional cylinder-DMRG studies, but now also have to include the interactions between the first and last column. While we only reported the ground state energies for the largest bond dimension $\chi=800$,  here we would like to provide more details. In particular, Table~\ref{tb:DMRG_data} shows the ground state energies per site for $\chi = 100,150,200,250,300,500,800$ and system sizes $L = 3,4,...,10$.}

\begin{figure}[htb]
    \includegraphics[width=1\columnwidth]{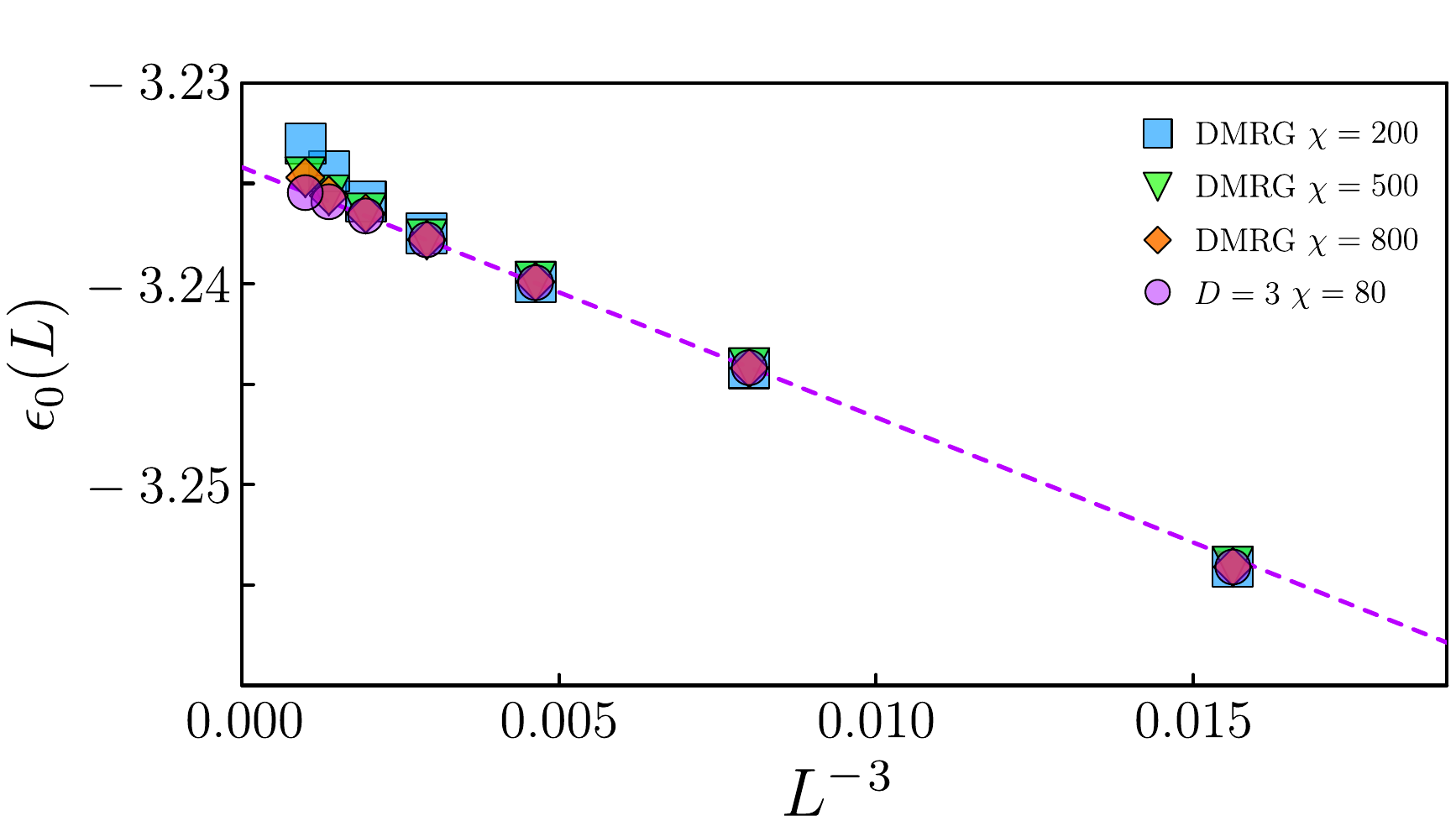}
\caption{Ground state energy $\epsilon_0$ at the critical point $\lambda_c = 3.04438$ as a function of linear system size $L$ from DMRG with truncation bond dimension $\chi=200,500,800$ compared with the energies obtained from PEPS with bond dimension $D=3$ following PTMRG approach. The dashed magenta line shows the energy scaling behavior for PEPS data points.}
    \label{fig:DMRG_energies}
\end{figure}

\changes{Fig.~\ref{fig:DMRG_energies} shows the comparison between DMRG energies and the ones obtained from PTMRG optimization for PEPS with local bond dimension $D=3$. One can see that while the PEPS results allow one to achieve robust scaling of the energies with the system size, the DMRG results show worse performance starting from the system size $L=8$. A similar conclusion can be drawn from Fig.~\ref{fig:mz_scaling} in the main text, where the magnetization $m_z$ obtained from DMRG cannot reproduce the desired scaling, while at the same time the proposed PEPS optimization provides a proper extrapolation of the scaling dimension.
}

\end{appendix}

\bibliography{main}

\end{document}